\documentclass[showpacs,amssymb,aps]{revtex4}

\usepackage{graphicx}
\begin{document}

\title{Path Integral Approach to Residual Gauge Fixing}  

\author{Ashok Das$^{a}$, J. Frenkel$^{b}$ and Silvana Perez$^{c}$}
\affiliation{$^{a}$ Department of Physics and Astronomy,
University of Rochester,
Rochester, New York 14627-0171, USA}
\affiliation{$^{b}$ Instituto de F\'{\i}sica, Universidade de S\~{a}o
Paulo, S\~{a}o Paulo, BRAZIL} 
\affiliation{$^{c}$ Departamento de F\'{\i}sica, 
Universidade Federal do Par\'{a}, 
Bel\'{e}m, Par\'{a} 66075-110, BRAZIL}

\bigskip

\begin{abstract}

In this paper we study the question of residual gauge fixing in the
path integral approach for a general class of axial-type gauges
including the light-cone gauge. We show that the two cases --
axial-type gauges and the light-cone gauge -- lead to very different
structures for the explicit forms of the propagator. In the case of
the axial-type gauges, fixing the residual symmetry determines the
propagator of the theory completely. On the other hand, in the
light-cone gauge there is still a prescription dependence even after
fixing the residual gauge symmetry, which is related to the existence
of an underlying global symmetry. 

\end{abstract}

\pacs{11.15.-q, 11.10.Ef, 12.38.Lg}

\maketitle

\section{Introduction}

Light-front field theories \cite{dirac} have been studied vigorously in the
past. Quantization on the light-front (as opposed to equal-time
quantization) leads to a larger number of kinematical generators of
the Poincar\'{e} algebra resulting in a trivial vacuum state of the
theory \cite{brodsky}.  Therefore, non-perturbative calculations can,
in principle, be carried out in a simpler manner in such theories.

More recently, it has been observed both in the conventional
light-front frame as well as in a generalized light-front frame
\cite{das,weldon,das1}  that
canonical quantization (within the Hamiltonian formalism) of a gauge
theory in the light-cone gauge
\begin{equation}
n\cdot A = 0,\qquad n^{2} = 0,\label{lightconegauge}
\end{equation}
where $A_{\mu}$ can be thought of as a matrix in the adjoint
representation of the gauge group in the case of a non-Abelian theory, leads
to a doubly transverse propagator \cite{brodsky1, perez}. For example,
in this case the
propagator has the explicit form (with the identity matrix neglected
in the case of a non-Abelian theory) 
\begin{equation}
D^{\rm (DT)}_{\mu\nu} (n,p) = - \frac{1}{p^{2}}\left[g_{\mu\nu} -
\frac{n_{\mu}p_{\nu} + n_{\nu}p_{\mu}}{(n\cdot p)} +
\frac{p^{2}}{(n\cdot
p)^{2}}\,n_{\mu}n_{\nu}\right],\label{doublytransverse}
\end{equation}
and satisfies
\begin{equation}
n^{\mu} D^{\rm (DT)}_{\mu\nu} (n,p) = 0 = D^{\rm (DT)}_{\mu\nu} (n,p)
n^{\nu} = p^{\mu} D^{\rm (DT)}_{\mu\nu} (n,p) = D^{\rm (DT)}_{\mu\nu}
(n,p) p^{\nu}.
\end{equation}

On the other hand, when calculated in the path integral formalism
using the naive Faddeev-Popov procedure, the
inverse of the two point function in the light-cone gauge
(\ref{lightconegauge}) takes the form (once again, we neglect the
identity matrix in the case of a non-Abelian theory)
\begin{equation}
\left(\Gamma^{\rm (PI)}\right)^{-1}_{\mu\nu} (n,p) = -
\frac{1}{p^{2}}\left[g_{\mu\nu} - \frac{n_{\mu}p_{\nu} +
n_{\nu}p_{\mu}}{(n\cdot p)}\right].\label{pathintegralinverse}
\end{equation}
While this is transverse with respect to $n^{\mu}$, it is not
transverse with respect to the momentum, namely,
\begin{eqnarray}
n^{\mu} \left(\Gamma^{\rm (PI)}\right)^{-1}_{\mu\nu} (n,p) & = & 0  =
\left(\Gamma^{\rm (PI)}\right)^{-1}_{\mu\nu} (n,p) n^{\nu},\nonumber\\
p^{\mu} \left(\Gamma^{\rm (PI)}\right)^{-1}_{\mu\nu} (n,p) & = &
\frac{n_{\nu}}{(n\cdot p)},\qquad \left(\Gamma^{\rm
(PI)}\right)^{-1}_{\mu\nu} (n,p) p^{\nu} = \frac{n_{\mu}}{(n\cdot p)}.
\end{eqnarray}
It is, of course, not clear {\em a priori} whether, in the light-cone
gauge, the inverse of the two point function corresponds exactly to
the propagator of the theory. We will show explicitly (in an appendix)
that such an identification can, in fact, be made in the light-cone
gauge. Therefore, there is a manifest difference
between the two structures in (\ref{doublytransverse}) and
(\ref{pathintegralinverse}) in
addition to the fact that one has to further specify a prescription
for handling the unphysical poles at $n\cdot p =0$. This difference 
has led to several papers \cite{suzuki} where the Lagrangian density
of the 
(Abelian) theory is modified arbitrarily by hand in order to reproduce
the propagator (\ref{doublytransverse}). 

We note that such a difference is not
restricted to light-front theories alone. In fact, even in a gauge
theory quantized at equal-time in the light-cone gauge, such a
difference does appear. Furthermore, even in the temporal gauge in
a theory quantized at equal-time, such a phenomenon arises. Normally,
one ascribes this to a residual gauge invariance in the path integral
approach. More specifically, in the canonical analysis of a gauge
theory, two first class constraints arise which necessitate two gauge
fixing conditions \cite{brodsky1,perez}, while in the naive path
integral approach one only imposes a single gauge fixing condition
such as in (\ref{lightconegauge}) which leaves behind a residual gauge
symmetry. However, a residual gauge
invariance has a very different effect in the sense that while the
original gauge invariance constrains the structure of the theory
strongly 
enough to make the two point function non-invertible, this is not the
case when there is a residual gauge invariance. Rather, in the
case of the temporal gauge, we know that residual gauge fixing removes
the arbitrary prescription dependence of the unphysical poles in the
propagator \cite{rossi, leibbrandt}. It is for this reason that we
would first like to understand the path integral formulation of the
theory in the light-cone gauge as much as is possible without fixing
the residual gauge
invariance before going into a systematic analysis of residual gauge
fixing. The paper is organized as follows. In section {\bf 2}, we
study systematically various properties of the theory in the naive
gauge fixing of the light-cone gauge in the path integral
formalism. We show, in particular, that the free theory in this gauge
has in general a global symmetry which allows for an arbitrary term
involving the tensor structure $n_{\mu}n_{\nu}$ in the
propagator. However, in this case, there arises an additional BRS
symmetry in the free action which restricts the path integral
propagator of the naively gauge fixed theory to have the form 
(\ref{pathintegralinverse}). The same arbitrariness of the tensor
structure in the propagator can also be understood as merely arising
as a result of a field redefinition which, in fact, is more along the
lines of a residual gauge fixing. In section {\bf 3}, we study the
question of fixing the residual gauge symmetry for the gauge $n\cdot A
= 0$ in a Yang-Mills theory for both axial-type gauges as well as the
light-cone 
gauge. The two cases have quite different features and so we discuss
them separately. We derive the forms of the completely gauge fixed
propagator in both the cases. In the case of axial-type
gauges the propagator has no further prescription dependence while the
unphysical poles in the 
propagator in the light-cone gauge do require a prescription even
after fixing the residual gauge symmetry. We trace 
the origin of such a behavior in the light-cone gauge to an underlying
global invariance of the free theory. We present a brief
conclusion in section {\bf 4}. In appendix {\bf A}, 
we show that the inverse of the two point function in the naive
light-cone gauge indeed corresponds to the propagator, while in
appendix {\bf B}, we compile some useful formulae for transformations
into light-cone variables.

\section{Naive Light-Cone Gauge Fixing}

In this section, we will study various properties of a light-cone
gauge fixed theory in the path integral formalism. Although the entire 
analysis in this paper will be carried out in the usual Minkowski
space-time for simplicity, all of our discussions
hold for a theory quantized either on a equal-time surface or on the 
light-front, both in the conventional as well as generalized
light-front frames (which are related by a change of
frame) \cite{brodsky,das,weldon,das1,perez}. Furthermore, since our
interest lies in studying the
structure of the propagator, we will restrict ourselves to analyzing
the free Maxwell theory in this section and comment on possible
differences which may arise in a fully interacting non-Abelian
theory. In the next section, our systematic analysis of residual gauge
fixing will be carried out within the context of a fully interacting
non-Abelian gauge theory.

The gauge invariant action for the Maxwell theory has the form
\begin{equation}
S_{\rm inv} = \int \mathrm{d}^{4}x\,{\cal L}_{\rm inv} =
-\frac{1}{4} \int \mathrm{d}^{4}x\,F_{\mu\nu}
F^{\mu\nu}.\label{action}
\end{equation}
The path integral  for the theory with a naive light-cone gauge fixing
(\ref{lightconegauge}) has the form \cite{faddeev}
\begin{equation}
Z = N \int {\cal D}A_{\mu}\,\Delta_{\rm FP}[A] \delta (n\cdot
A)\,e^{iS_{\rm inv}},\label{generatingfunction}
\end{equation}
where $N$ is a normalization constant and  $\Delta_{\rm FP}[A]$
represents the Faddeev-Popov determinant
which can, in general, be field dependent in a non-Abelian
theory. However, in a general axial-type gauge (including the
light-cone gauge 
that we are analyzing here), it is known that the Faddeev-Popov
determinant is independent of the field variables even in a
non-Abelian theory \cite{frenkel}. 

It is clear from the structure of the generating functional in
(\ref{generatingfunction}) that any $n$-point Green's function
involving the gauge fields will be
transverse with respect to the light-like vector $n^{\mu}$, namely,
\begin{equation}
n^{\mu_{1}} \langle 0|T\left(A_{\mu_{1}}\cdots
A_{\mu_{n}}\right)|0\rangle = N \int {\cal D}A_{\mu}\,\Delta_{\rm FP}
[A] \delta (n\cdot A) n\cdot A A_{\mu_{2}}\cdots
A_{\mu_{n}}\,e^{iS_{\rm inv}} = 0.
\end{equation}
In particular, this would imply that the propagator would satisfy
\begin{equation}
n^{\mu} D_{\mu\nu} (n,p) = 0,\label{ntransverse}
\end{equation}
which is satisfied by both (\ref{doublytransverse}) and
(\ref{pathintegralinverse}). 

We note that when $n^{2}=0$, the gauge
fixing condition as well as the invariant action
$S_{\rm inv}$ in (\ref{action}) are invariant under the
infinitesimal global transformation 
\begin{equation}
\delta A_{\mu} = \zeta n_{\mu}\,\partial\cdot A,\label{globaltfn}
\end{equation}
where $\zeta$ represents the constant infinitesimal parameter of
transformation. The change in the path integral measure under such a field
redefinition is easily seen to be a field independent
constant. Conversely, if we incorporate the gauge fixing
condition as well as the Faddeev-Popov determinant into the action, we
can write the generating functional in the form
\begin{equation}
Z = N \int {\cal D}A_{\mu} {\cal D}F {\cal D}\overline{c} {\cal D}
c\,e^{iS},
\end{equation}
where
\begin{equation}
S = S_{\rm inv} - \int \mathrm{d}^{4}x\left( F n\cdot A -
\overline{c} n\cdot \partial c\right), 
\end{equation}
and $F$ represents an auxiliary field implementing the gauge
condition. It is easy to verify that the  action as well as 
the generating functional in this case are invariant
under the infinitesimal global transformations
\begin{equation}
\delta A_{\mu} = \zeta n_{\mu} \partial\cdot A,\qquad \delta F =
\zeta \Box\,\partial\cdot A,\qquad \delta c = 0 = \delta
\overline{c}.\label{globaltfn1}
\end{equation}

If we add sources into the path integral as
\begin{equation}
Z [J^{\mu}, J, \eta, \overline{\eta}] = e^{i W [J^{\mu}, J, \eta,
\overline{\eta}]} = N \int {\cal D} A_{\mu} {\cal D} F {\cal D}
\overline{c} {\cal D} c\,e^{i\left(S + S_{\rm source}\right)},\qquad
S_{\rm source} = \int \mathrm{d}^{4}x\left(
J^{\mu}A_{\mu} + JF + i \left(\overline{\eta} c -
\overline{c} \eta\right)\right),\label{generatingfunction1}
\end{equation}
then, the Ward identity for the global invariance of
(\ref{globaltfn1}) can be easily derived to be
\begin{equation}
\int \mathrm{d}^{4}x\,\left(n\cdot J (x) \partial_{\mu} \frac{\delta
W}{\delta J_{\mu} (x)} + J (x) \Box\,\partial_{\mu} \frac{\delta
W}{\delta J_{\mu} (x)}\right) = 0.\label{wi}
\end{equation}
This leads to a constraint on the form of the gauge propagator of the
form
\begin{equation}
n^{\nu} \partial^{(x)}_{\mu} \frac{\delta^{2} W}{\delta J_{\mu} (x)
\delta J_{\lambda} (y)} + n^{\lambda} \partial^{(y)}_{\mu}
\frac{\delta^{2} W}{\delta J_{\mu} (y) \delta J_{\nu} (x)} = 0.
\end{equation}
Together with (\ref{ntransverse}), this determines the general form of
the propagator for the gauge field in the light-cone gauge
(\ref{lightconegauge}) to be 
\begin{equation}
D_{\mu\nu} (n,p) = - \frac{1}{p^{2}}\left[\eta_{\mu\nu} -
\frac{n_{\mu}p_{\nu} + n_{\nu} p_{\mu}}{(n\cdot p)} +
\alpha\,\frac{p^{2}}{(n\cdot
p)^{2}}\,n_{\mu}n_{\nu}\right],\label{generalpropagator}
\end{equation}
where $\alpha$ is arbitrary. We note that when $n^{2} = 0$, this
satisfies $n^{\mu} D_{\mu\nu} (n,p) = 0$ as is expected from
(\ref{ntransverse}). For $\alpha = 1$, this corresponds to the doubly
transverse propagator of (\ref{doublytransverse}) while for $\alpha =
0$, this coincides with the path integral propagator of
(\ref{pathintegralinverse}).  It is worth noting here that if we
identify the tensor structure in (\ref{generalpropagator}) with the
sum over polarization vectors,
\begin{equation}
\sum_{\lambda} \epsilon_{\mu} (p,\lambda) \epsilon_{\nu} (p,\lambda) =
- \eta_{\mu\nu} + \frac{n_{\mu}p_{\nu} + n_{\nu} p_{\mu}}{(n\cdot p)} -
\alpha\,\frac{p^{2}}{(n\cdot
p)^{2}}\,n_{\mu}n_{\nu},\label{polarizationsum} 
\end{equation}
where we have chosen the polarization vector $\epsilon_{\mu}
(p,\lambda)$ to be real, then, when $n^{2}=0$, we see that for any
value of $\alpha$, we have
\begin{equation}
\sum_{\lambda} \epsilon^{\mu} (p,\lambda) \epsilon_{\mu} (p,\lambda) =
- 2.
\end{equation}
This does not, however, imply that the polarization vectors are summed
over only the physical ones in (\ref{polarizationsum}). In fact, the
arbitrariness in the $n_{\mu}n_{\nu}$ term in (\ref{polarizationsum})
signifies that the polarization sum does contain polarization vectors
proportional to the light-like vector, $\epsilon_{\mu} (p,\lambda)
\sim n_{\mu}$. On the other hand, such polarization vectors would lead
to zero norm states and, therefore, cannot represent physical
polarizations.

This issue can be further understood by noting that while 
\begin{equation}
P_{\mu\nu}^{\rm (T)} = \eta_{\mu\nu} -
\frac{n_{\mu}n_{\nu}}{n^{2}},\qquad P_{\mu\nu}^{\rm (L)} =
\frac{n_{\mu}n_{\nu}}{n^{2}},
\end{equation}
define transverse and longitudinal projection operators with respect
to a vector $n^{\mu}$ when $n^{2}\neq 0$, they are not defined for a
light-like vector. In fact, one can define transverse and longitudinal
projection operators for a light-like vector only in conjunction with
another vector with a nonzero inner product. Thus, for example, if we
take the second vector as the gradient operator, then for a light-like
vector $n^{\mu}$ we can define \cite{perez}
\begin{equation}
P_{\mu\nu}^{\rm (T)} (n,\partial) = \eta_{\mu\nu} -
\frac{n_{\mu}\partial_{\nu} + n_{\nu}\partial_{\mu}}{(n\cdot
\partial)} + \frac{\partial^{2}}{(n\cdot \partial)^{2}}\,n_{\mu}
n_{\nu}, \qquad
P_{\mu\nu}^{\rm (L)} = \frac{n_{\mu}\partial_{\nu} +
n_{\nu}\partial_{\mu}}{(n\cdot \partial)} -
\frac{\partial^{2}}{(n\cdot
\partial)^{2}}\,n_{\mu}n_{\nu}.\label{projection}
\end{equation}
It can be checked that these define orthogonal projection operators
and satisfy
\begin{eqnarray}
n^{\mu} P_{\mu\nu}^{\rm (T)} & = & 0 = P_{\mu\nu}^{\rm (T)}
n^{\nu} = \partial^{\mu} P_{\mu\nu}^{\rm (\rm T)} = P_{\mu\nu}^{\rm
(T)} \partial^{\nu},\nonumber\\
n^{\mu} P_{\mu\nu}^{\rm (L)} & = & n_{\nu},\qquad P_{\mu\nu}^{\rm (L)}
n^{\nu} = n_{\mu},\qquad \partial^{\mu} P_{\mu\nu}^{\rm (L)} =
\partial_{\nu},\qquad P_{\mu\nu}^{\rm (L)} \partial^{\nu} =
\partial_{\mu},\nonumber\\
P_{\mu\nu}^{\rm (T)} + P_{\mu\nu}^{\rm (L)} & = &
\eta_{\mu\nu}.\label{relations} 
\end{eqnarray}
This allows us to decompose any vector and, in particular, the gauge
field as
\begin{equation}
A_{\mu} = A_{\mu}^{\rm (T)} + A_{\mu}^{\rm (L)},\qquad A_{\mu}^{\rm
(T)} = P_{\mu\nu}^{\rm (T)} A^{\nu},\quad A_{\mu}^{\rm (L)} =
P_{\mu\nu}^{\rm (L)}A^{\nu}.\label{decomposition}
\end{equation}
We note that by construction,
\begin{equation}
n\cdot A^{\rm (T)} = 0 = \partial\cdot A^{\rm (T)},\qquad n\cdot
A^{\rm (L)} = n\cdot A,\qquad \partial\cdot A^{\rm (L)} =
\partial\cdot A.
\end{equation}
Therefore, each of the four vectors $A_{\mu}^{\rm (T)}, A_{\mu}^{\rm
(L)}$ carries only two degrees of freedom. Furthermore, under a gauge
transformation,
\begin{equation}
\delta A_{\mu} = \partial_{\mu}\theta (x),
\end{equation}
it follows, using (\ref{relations}), that
\begin{equation}
\delta A_{\mu}^{\rm (T)} = 0,\qquad \delta A_{\mu}^{\rm (L)} =
\partial_{\mu} \theta (x).
\end{equation}
Consequently, we see that $A_{\mu}^{\rm (T)}$ is gauge invariant and
carries only the physical degrees of freedom while $A_{\mu}^{\rm (L)}$
consists of the two unphysical (gauge) degrees of freedom. From the
definition of $A_{\mu}^{\rm (L)}$ in (\ref{decomposition}) and
(\ref{projection}), we see that it is completely determined from a
knowledge of $n\cdot A$ and $\partial\cdot A$. While the gauge fixing
condition (\ref{lightconegauge}) specifies one of the components,
$\partial\cdot A$ remains arbitrary and the transformation
(\ref{globaltfn}) (or (\ref{globaltfn1})) merely reflects the
arbitrariness in this component.

So far we have discussed the consequences on the structure of the
propagator following from the global invariance in
(\ref{globaltfn}). Such a global invariance will be quite important in
the analysis of the form of the propagator after residual gauge
fixing to be discussed in the 
next section. The Ward identity (\ref{wi}) is quite general and
does not depend on the particular structure of the theory. For
example, if the action had an additional term of the form
\begin{equation}
S_{\rm additional} = - \frac{1}{2\xi} \int
\mathrm{d}^{4}x\,\left(\partial\cdot A\right)^{2},
\end{equation}
it would still be invariant under the global transformation
(\ref{globaltfn}) and the Ward identity (\ref{wi}) would continue to
hold. Thus, it is quite curious as to why the path integral propagator
has a unique form corresponding to $\alpha = 0$. This constraint, in
fact, comes from an additional BRS invariance that the free theory
develops. In fact, it can be checked that in addition to the usual BRS  
transformations \cite{BRS} under which the action
(\ref{generatingfunction1})  
is invariant, it is also invariant under a new BRS transformation of
the form
\begin{equation}
\tilde{\delta} A_{\mu} = \tilde{\omega} n_{\mu} c,\qquad
\tilde{\delta}c = 0,\qquad \tilde{\delta}\overline{c} = \tilde{\omega}
\partial\cdot A,\qquad \tilde{\delta} F = \tilde{\omega} \Box
c,\label{newbrs} 
\end{equation}
where $\tilde{\omega}$ represents an anti-commuting global parameter. This
transformation anti-commutes with the usual BRS transformation and is
also nilpotent (like the conventional BRS transformation), but only on-shell
(when the ghost equations of motion are used) which is a reflection of
the absence of some auxiliary field in the theory. This new BRS
invariance leads to a Ward identity of the form
\begin{equation}
\int \mathrm{d}^{4}x\,\left[\partial_{\mu} \frac{\delta W}{\delta J_{\mu}
(x)}\,\eta (x) + n\cdot J (x)\,\frac{\delta W}{\delta \overline{\eta}
(x)} + J (x) \Box \frac{\delta W}{\delta \overline{\eta} (x)}\right] =
0.\label{wi1}
\end{equation}
It is clear that unlike the global transformation (\ref{globaltfn}),
the BRS invariance is very specific to a specific theory and,
correspondingly, the resulting Ward identity is also. From
(\ref{wi1}), it can be easily derived that
\begin{equation}
\partial^{(x)}_{\mu} \frac{\delta^{2} W}{\delta J_{\mu} (x) \delta
J_{\nu} (y)} - n^{\nu}\,\frac{\delta^{2} W}{\delta \overline{\eta}
(y)\delta \eta (x)} = 0.
\end{equation}
This relates the divergence of the gauge propagator to the ghost
propagator (it does not say that the gauge propagator is transverse)
and thereby determines $\alpha = 0$.

It is worth pointing out here that the global invariance of
(\ref{globaltfn}) or the new BRS invariance of (\ref{newbrs}) cannot
be incorporated into a fully interacting theory which would include a
non-Abelian theory. However, the violation of the invariance will
occur only at the higher order terms in the number of fields. Namely,
the violation of the Ward identities in (\ref{wi}) or (\ref{wi1}) will
manifest only in the structure of the higher point functions. As far
as the structure of the propagator is concerned, all of our
discussions will hold in a fully interacting theory as well.

There is yet another interesting and suggestive way to see the
arbitrariness in the  $n_{\mu}n_{\nu}$ term in the propagator. For
example suppose we start with the path integral
propagator in (\ref{pathintegralinverse}) in the momentum space, namely,
\begin{equation}
\langle 0|T \left(A_{\mu} (x) A_{\nu} (y)\right)|0\rangle \rightarrow
- \frac{1}{p^{2}}\left[\eta_{\mu\nu} - \frac{n_{\mu}p_{\nu} + n_{\nu}
p_{\mu}}{(n\cdot p)}\right].  
\end{equation}
Then, under a field redefinition
\begin{equation}
A'_{\mu} (x) = A_{\mu} (x) - \frac{1}{2\beta}\,n_{\mu} \frac{1}{n\cdot
\partial}\,\partial\cdot A,\label{redefinition}
\end{equation}
which preserves the gauge fixing condition, the propagator would
change in the momentum space to ($\beta$ is a constant)
\begin{equation}
\langle 0|T\left(A'_{\mu} (x) A'_{\nu} (y)\right)|0\rangle \rightarrow
- \frac{1}{p^{2}}\left[\eta_{\mu\nu} - \frac{n_{\mu}p_{\nu} + n_{\nu}
p_{\mu}}{(n\cdot p)} + \alpha\,\frac{p^{2}}{(n\cdot
p)^{2}}\,n_{\mu}n_{\nu}\right],
\end{equation}
where we have identified $\alpha = \frac{4\beta -1}{4\beta^{2}}$. This
has the same structure as (\ref{generalpropagator}). This derivation,
however, is quite suggestive for the following reason. Let us consider a
field redefinition of the form (\ref{redefinition}) in the path
integral (\ref{generatingfunction}). While the gauge fixing condition
is invariant under such a redefinition, the action is not. In fact,
under this redefinition
\begin{equation}
S_{\rm inv} [A] = S_{\rm inv} [A'] - \frac{1}{2\xi} \int
\mathrm{d}^{4}x\,\left(\partial\cdot
A'\right)^{2},\label{covariantgaugefixing} 
\end{equation}
where we have identified $\xi = \frac{(2\beta
-1)^{2}}{1-4\beta}$. This shows that the field redefinition induces an
additional term in the action which is reminiscent of a covariant
gauge fixing term. In fact, it can be written in the path integral in
the form of a delta function $\delta (\partial\cdot A' -
\sqrt{\xi} f)$. While this is suggestive and seems to imply that the
field redefinition must somehow correspond to fixing the residual
gauge invariance of the theory, it is not quite complete for a variety
of reasons. First, the field redefinition is meaningful only over a
limited range of the parameters $\xi$. In fact, the Jacobian of the
field transformation which has the form
\begin{equation}
J = \det\left|\frac{\partial A_{\mu}}{\partial A'_{\nu}}\right| =
\frac{2\beta}{2\beta -1},
\end{equation}
becomes singular for $\beta = \frac{1}{2}$ exactly at the point where
$\xi = 0$. Furthermore, we do not quite see the Faddeev-Popov
determinant arising from such a field redefinition. Thus, we conclude
from all this analysis that a systematic understanding of the residual
gauge fixing 
is necessary in order to fully appreciate the structure of the
propagator in the path integral formalism and we will do this in the
next section.

\section{Residual Gauge Fixing}

The question of residual gauge fixing within the path integral
approach for the temporal gauge has
previously been studied in some detail in \cite{rossi,girotti}.
In this section, we will systematically study the question of gauge
fixing for the residual gauge invariance in a Yang-Mills theory in a 
general class of axial-type gauges including the light-cone
gauge. However, we will divide the study into two cases -- 
axial-type gauges and the light-cone gauge -- because as we will see the
two cases lead to quite different results. 

\subsection{Axial-type Gauges}

Let us consider a Yang-Mills theory described by the gauge invariant
action 
\begin{equation}
S_{\rm inv} = - \frac{1}{4}\int \mathrm{d}^{4}x\,{\rm
Tr}\,F_{\mu\nu}F^{\mu\nu},\label{action1}
\end{equation}
where the gauge fields are assumed to belong to the adjoint
representation of the gauge group and the field strength is defined to
be 
\begin{equation}
F_{\mu\nu} = \partial_{\mu}A_{\nu} - \partial_{\nu} A_{\mu} - i
\left[A_{\mu},A_{\nu}\right].\label{fieldstrength}
\end{equation}
For simplicity, we have scaled the coupling constant to unity. The
action in (\ref{action1}) is invariant under the infinitesimal gauge
transformation
\begin{equation}
\delta A_{\mu} = D_{\mu}\theta (x) = \partial_{\mu}\theta (x) - i
\left[A_{\mu} (x),\theta (x)\right],\label{gaugetfn}
\end{equation}
where $\theta (x)$ is the infinitesimal parameter of transformation.

Let us consider an axial-type gauge of the form
\begin{equation}
n\cdot A = 0,\qquad |n^{2}| = 1.\label{axialgauge}
\end{equation}
This can, therefore, describe the temporal gauge or the axial gauge
depending on the choice of the vector $n^{\mu}$. Furthermore, since
the gauge condition (\ref{axialgauge}) involves a Lorentz scalar, it
can hold in any frame including the general light-front frame (which
is not related to the Minkowski frame through a Lorentz
transformation, for details see \cite{perez}). 
The naive generating functional
\begin{equation}
Z = N \int {\cal D} A_{\mu}\,e^{iS_{\rm inv}},\label{naivegf}
\end{equation}
does not exist because of the gauge invariance of the
theory. We would like to impose the general axial gauge
(\ref{axialgauge}) and following Faddeev-Popov \cite{faddeev}, we
introduce the identity
\begin{equation}
\Delta_{\rm FP}^{-1} [A] = \int {\cal D}\theta\,\delta \left(n\cdot
A^{(\theta)}\right),\label{fp1}
\end{equation}
where $A_{\mu}^{(\theta)}$ represents a gauge transformed potential
(see (\ref{gaugetfn}))
\begin{equation}
A_{\mu}^{(\theta)} (x) = A_{\mu} (x) + D_{\mu}\theta (x).
\end{equation}
The Faddeev-Popov determinant in (\ref{fp1}) is manifestly gauge
invariant and the standard procedure of Faddeev-Popov can be followed
to separate the volume of gauge orbits as
\begin{eqnarray}
Z & = & N \int {\cal D}A_{\mu} \Delta_{\rm FP}[A]\int {\cal
D}\theta\,\delta \left(n\cdot A^{(\theta)}\right)\,e^{iS_{\rm
inv}}\nonumber\\
 & = & N \left(\int {\cal D}\theta\right) \int {\cal D}
A_{\mu}\,\Delta_{\rm FP} [A]\delta (n\cdot A)\,e^{iS_{\rm
inv}}.\label{generatingfunction2}
\end{eqnarray}
In the derivation above, we have inserted the identity from
(\ref{fp1}) in the first step while we have made an inverse gauge
transformation and used the gauge invariance of the Faddeev-Popov
determinant in the second step. This would, therefore, seem to have
separated out the infinite gauge volume element from the path
integral. However, this is not entirely true.

In fact, let us note that the delta function constraint in
(\ref{generatingfunction2}) is invariant under gauge transformations
of the form (the Faddeev-Popov determinant is gauge invariant) 
\begin{equation}
A_{\mu} (x) \rightarrow A_{\mu} (x) + D_{\mu}\overline{\theta}
(x),\qquad n\cdot \partial \overline{\theta} (x) =
0,\label{restricted}
\end{equation}
where $\overline{\theta} (x)$ is an arbitrary function independent of
$n\cdot x$.
Namely, in the case of axial-type gauges, the gauge transformation
parameters can be grouped into two classes -- ones that do not depend on
the coordinate $n\cdot x$ and others that do -- and the gauge fixing
condition (\ref{axialgauge}) cannot determine the transformation
parameters which do not depend on the coordinate $n\cdot x$. This is
the reason why the separation of the infinite gauge volume is
incomplete in (\ref{generatingfunction2}) and manifests in a residual
gauge symmetry of the generating functional since each factor in the
path integral is invariant under gauge transformations of the form
(\ref{restricted}).  

To separate out the volume associated with the residual (restricted)
gauge 
transformations, we will follow again the method of Faddeev-Popov and
fix a gauge. We note that a covariant gauge condition such as
\begin{equation}
\partial\cdot A = \sqrt{\xi} f (x),\label{residual}
\end{equation}
where $\xi$ is an arbitrary constant and $f (x)$ is an arbitrary
function,  can be implemented through a gauge transformation of the
type (\ref{restricted}) provided
\begin{equation}
\overline{\theta} (x) = \frac{1}{\partial\cdot D} \left(-\partial\cdot
A (x) + \sqrt{\xi} f (x)\right).\label{restricted1}
\end{equation}
However, since $\overline{\theta} (x)$ does not depend on $n\cdot x$,
such a condition (\ref{restricted1}) makes sense only at a given value of
$n\cdot x = \tau$ where $\tau$ is an arbitrary fixed constant. Thus, the
residual gauge fixing condition
(\ref{residual}) can be implemented only at a fixed $n\cdot x = \tau$. 
In this case, we can use the identity,
\begin{equation}
\left.\bar{\Delta}_{\rm FP}^{-1} [A]\right|_{n\cdot x=\tau} =
\int {\cal D}\overline{\theta}\,\delta \left(\partial\cdot
A^{(\overline{\theta})} - \sqrt{\xi} f(x)\right)_{n\cdot
x=\tau}.\label{fp2}
\end{equation}
This second Faddeev-Popov
determinant in (\ref{fp2}) is manifestly invariant under a restricted
gauge transformation (\ref{restricted}) and is defined on the space of
functions annihilated by $n\cdot \partial$ (see
(\ref{restricted})). Following the earlier derivation in
(\ref{generatingfunction2}), we can now write the generating function
as  
\begin{eqnarray}
Z & = & N \left(\int {\cal D}\theta\right)\int {\cal D}
A_{\mu}\,\Delta_{\rm FP} [A] \delta (n\cdot A)\,\left.\bar{\Delta}_{\rm
FP} [A]\right|_{n\cdot x=\tau} \int {\cal D}\overline{\theta}\,\delta
\left(\partial\cdot
A^{(\overline{\theta})} - \sqrt{\xi} f (x)\right)_{n\cdot
x=\tau}\,e^{iS_{\rm inv}}\nonumber\\
 & = & N \left(\int {\cal D}\theta\right)\left(\int {\cal
D}\overline{\theta}\right) \int {\cal D} A_{\mu}\,\Delta_{\rm FP}
[A]\left.\bar{\Delta}_{\rm FP} [A]\right|_{n\cdot x=\tau}
\delta (n\cdot A) \delta
\left(\partial\cdot A - \sqrt{\xi} f\right)_{n\cdot x=\tau}\,e^{iS_{\rm
inv}}.\label{generatingfunction3}
\end{eqnarray}
The gauge volume is now completely extracted and can be absorbed into
the normalization factor $N$. 

The second delta function can be
exponentiated using the 't Hooft trick of using a
Gaussian weight factor \cite{thooft} leading to
\begin{equation}
S = S_{\rm inv} - \frac{1}{2\xi} \int \mathrm{d}^{4}x\,\delta (n\cdot
x-\tau)\left(\partial\cdot A\right)^{2},\label{secondfixing}
\end{equation}
which resembles (\ref{covariantgaugefixing}) except for the fact that
it is defined only for a fixed value of $n\cdot x$. The generating
functional takes the form
\begin{equation}
Z = N \int {\cal D} A_{\mu}\,\Delta_{\rm FP} [A]
\left.\bar{\Delta}_{\rm FP} [A]\right|_{n\cdot x=\tau} \delta (n\cdot
A)\,e^{iS},\label{generatingfunction4}
\end{equation}
where we have absorbed the gauge volume elements into the
normalization constant. As is well known, in an axial-type gauge, the
Faddeev-Popov determinant $\Delta_{\rm FP} [A]$ is trivial
\cite{frenkel}. However,
the determinant coming from the second gauge fixing is not and can be
written in the form of an action of the form
\begin{equation}
\left.\bar{\Delta}_{\rm FP} [A]\right|_{n\cdot x=\tau} = \int {\cal
D}\overline{c}\,{\cal D}c\,e^{iS_{\rm ghost}},\qquad S_{\rm ghost} =
- \int \mathrm{d}^{4}x\,\delta (n\cdot x - \tau)\overline{c}
\partial^{\mu} D_{\mu}c.\label{ghostaction}
\end{equation}

To determine the propagator, we note that in the axial-type gauges
(\ref{axialgauge}) with the second gauge fixing
term (\ref{secondfixing}) in the action, the Green's function for the
theory has to satisfy the equation
\begin{equation}
\left(\delta^{\mu}_{\sigma} - \frac{n^{\mu}n_{\sigma}}{n^{2}}\right)
\left(\eta^{\sigma\lambda} \Box - \partial^{\sigma}\partial^{\lambda} +
\frac{1}{\xi} \delta (n\cdot x - \tau)
\partial^{\sigma}\partial^{\lambda}\right) D_{\lambda\nu} (x,y) =
\left(\delta^{\mu}_{\nu} - \frac{n^{\mu}n_{\nu}}{n^{2}}\right) 
\delta^{4} (x-y),\quad n^{\mu} D_{\mu\nu} = 0 =
D_{\mu\nu}n^{\nu}.\label{greensfunction}
\end{equation}
We note that with the available tensor structures, we can construct
two linearly independent, orthogonal second rank symmetric projection
operators 
which will be transverse to the vector $n^{\mu}$. Namely,
\begin{eqnarray}
P_{\mu\nu} & = & \eta_{\mu\nu} - \epsilon (n^{2}) n_{\mu}n_{\nu} -
\frac{\left(\partial_{\mu} - \epsilon (n^{2}) n_{\mu} n\cdot
\partial\right)\left(\partial_{\nu} - \epsilon (n^{2}) n_{\nu} n\cdot
\partial\right)}{\left(\Box - \epsilon (n^{2}) (n\cdot
\partial)^{2}\right)}\nonumber\\
Q_{\mu\nu} & = & \frac{\left(\partial_{\mu} - \epsilon (n^{2}) n_{\mu} n\cdot
\partial\right)\left(\partial_{\nu} - \epsilon (n^{2}) n_{\nu} n\cdot
\partial\right)}{\left(\Box - \epsilon (n^{2})(n\cdot
\partial)^{2}\right)}, 
\end{eqnarray}
where $\epsilon (n^{2})$ represents the sign of $n^{2}$. Each of these
two structures satisfies
\begin{equation}
n^{\mu} P_{\mu\nu} = 0 = n^{\mu}Q_{\mu\nu} = P_{\mu\nu} n^{\nu} =
Q_{\mu\nu} n^{\nu}.
\end{equation}
However, it is easy to check that the first structure, in addition, is
transverse to $\partial^{\mu}$,
\begin{equation}
\partial^{\mu} P_{\mu\nu} = 0 = P_{\mu\nu} \partial^{\nu}.
\end{equation}

Since the propagator has to be transverse to $n^{\mu}$, we can expand
it as
\begin{equation}
D_{\mu\nu} (x,y) = P_{\mu\nu} a (x,y) + Q_{\mu\nu}
b(x,y),\label{propagatordecomposition}
\end{equation}
where we can think of $a(x,y), b(x,y)$ respectively as the transverse
and the longitudinal components of the propagator. Substituting
(\ref{propagatordecomposition}) into (\ref{greensfunction}), it is
easily determined that $a (x,y)$ and $b (x,y)$ satisfy
\begin{equation}
\Box a (x,y) = \delta^{4} (x-y),\qquad \left(\epsilon (n^{2})\left(n\cdot
\partial\right)^{2} + \frac{1}{\xi} \delta (n\cdot x - \tau)
\left(\Box - \epsilon (n^{2})\left(n\cdot
\partial\right)^{2}\right)\right) b (x,y) =
\delta^{4} (x-y).\label{equations}
\end{equation}
The first equation is straightforward to solve and gives (in four
dimensions) 
\begin{equation}
a (x,y) = - \int \frac{\mathrm{d}^{4}k}{(2\pi)^{4}}\,
\frac{e^{ik\cdot (x-y)}}{k^{2}}.\label{transverse}
\end{equation}
The second equation in (\ref{equations}) is a bit more involved
because of the delta function, but leads to a solution of the form
\begin{equation}
b (x,y) = \int 
\frac{\mathrm{d}^{3}k_{T}}{(2\pi)^{3}}\,e^{-ik_{T}\cdot
(x_{T}-y_{T})}\left[-\frac{\xi}{k_{T}^{2}} + \epsilon (n^{2})\left(
  |n\cdot x- n\cdot y| - |n\cdot x - \tau| - |n\cdot y -
  \tau|\right)\right]. \label{longitudinal}
\end{equation}
Here we have defined the transverse coordinates and momenta as
\begin{equation}
x^{\mu}_{T} = x^{\mu} - \epsilon (n^{2}) n^{\mu} (n\cdot x),\qquad
k_{\mu}^{T} = k_{\mu} - \epsilon (n^{2}) n_{\mu} (n\cdot k).
\end{equation}

This defines the completely gauge fixed propagator which is well
behaved without 
any unphysical pole for any finite value of $\xi$. In particular, for
$\xi=0$, we note that the longitudinal part of the propagator
(\ref{longitudinal})  vanishes for $n\cdot x = \tau$ or $n\cdot y =
\tau$ as we would expect from the residual gauge fixing. In the
temporal gauge, for example, $n^{\mu} = (1,0,0,0)$ and with the
transverse and the longitudinal parts given in (\ref{transverse}) and
(\ref{longitudinal}) respectively, the propagator
(\ref{propagatordecomposition}) takes the form
\begin{equation}
D_{ij} (x,y) = - \int
\frac{\mathrm{d}^{4}k}{(2\pi)^{4}}
\left(\eta_{ij} +
\frac{k_{i}k_{j}}{\vec{k}^{2}}\right)\frac{e^{ik\cdot (x-y)}}{k^{2}} -
\int
\frac{\mathrm{d}^{3}k}{(2\pi)^{3}}\,\frac{k_{i}k_{j}}{\vec{k}^{2}}\,
e^{-i\vec{k}\cdot (\vec{x}-\vec{y})}\left(\frac{\xi}{\vec{k}^{2}} +
|x^{0}-y^{0}| - |x^{0}-\tau| -
|y^{0}-\tau|\right).\label{temporalgauge} 
\end{equation}
For $\xi = 0$, the form of the propagator in the temporal gauge in
(\ref{temporalgauge}) had 
already been obtained in \cite{rossi} where it has also been argued that the
longitudinal part of the propagator is quite crucial in obtaining the
correct value for the Wilson line.
It is worth recalling that the propagator for the theory in the path
integral in the temporal gauge (without the residual gauge fixing) has
the form
\begin{eqnarray}
D_{ij} (x,y) & = & - \int
\frac{\mathrm{d}^{4}k}{(2\pi)^{4}}\,\frac{e^{ik\cdot (x-y)}}{k^{2}}
\left(\eta_{ij} + \frac{k_{i}k_{j}}{k_{0}^{2}}\right)\nonumber\\
 & = & - \int \frac{\mathrm{d}^{4}k}{(2\pi)^{4}}\,e^{ik\cdot
(x-y)}\left[\frac{1}{k^{2}}\left(\eta_{ij}+
\frac{k_{i}k_{j}}{\vec{k}^{2}}\right) -
\frac{1}{k_{0}^{2}}\,\frac{k_{i}k_{j}}{\vec{k}^{2}}\right].
\end{eqnarray}
We note from the above expression that while the transverse part of
the propagator is well defined without any unphysical pole and
coincides with  that in (\ref{temporalgauge}), the longitudinal part
depends on the prescription for handling the pole at $k_{0}=0$. The
residual gauge fixing has removed this arbitrariness in the
longitudinal part in (\ref{temporalgauge}) for any finite value of
$\xi$. This is, in fact, a very general result. As we have argued in
the last section, any residual invariance of the quadratic part of the
action 
-- local or global -- does lead to an arbitrariness in the definition
of the propagator which reflects in some form of prescription
dependence. In the case of axial-type gauges, this is completely fixed
by the residual gauge fixing. However, as we will show next, the
behavior is quite different in the light-cone gauge.  

\subsection{Light-cone Gauge}

Let us next study the action (\ref{action1}) in the light-cone gauge
\begin{equation}
n\cdot A = 0,\qquad n^{2} = 0,\label{lightconegauge1}
\end{equation}
which has been studied by various groups from different points of view
\cite{leibbrandt,bassetto}.
The discussion for the gauge fixing in the path integral approach, in
this case, follows exactly as discussed for the axial-type gauges and
we obtain the generating functional in (\ref{generatingfunction2}) as
a result of the naive light-cone gauge fixing. The analysis of the
residual symmetry, however, differs from the earlier case.

Let us note that given a light-like vector $n^{\mu}$, one can define a
dual light-like vector $\tilde{n}^{\mu}$ such that
\begin{equation}
\tilde{n}^{2} = 0,\qquad n\cdot \tilde{n} \neq 0.
\end{equation}
For example, with $n^{\mu} = (1,0,0,-1)$, we can define $\tilde{n}^{\mu}
= (1,0,0,1)$. Correspondingly, one can label the coordinates as
$x^{\mu} = (n\cdot x, \tilde{n}\cdot x, x^{\mu}_{T})$ where
$x^{\mu}_{T}$ is transverse to both $n^{\mu}$ and
$\tilde{n}^{\mu}$. In such a case, the delta function constraint in
(\ref{generatingfunction2}) can again be seen to be invariant under a
residual gauge transformation (\ref{restricted})
\begin{equation}
A_{\mu} (x) \rightarrow A_{\mu} (x) + D_{\mu} \overline{\theta}
(x),\qquad n\cdot\partial \overline{\theta} (x) = 0. \label{restricted2}
\end{equation}
However, because of the light-like nature of $n^{\mu}$, the
implications of (\ref{restricted2}) in this case are different and,
in particular, it implies that the parameter of gauge transformation
must be independent of $\tilde{n}\cdot x$. This difference from the
earlier case leads to the essential difference in the structure of the
propagator in the light-cone gauge.

Once again, as in the axial-type gauges, if we would like to impose a
covariant gauge fixing, for the residual gauge symmetry, of the form
\begin{equation}
\partial\cdot A (x) = \sqrt{\xi} f (x),
\end{equation}
we can implement it only at a fixed value of $\tilde{n}\cdot x=\tau$
since the parameter of the residual gauge transformation does not
depend on $\tilde{n}\cdot x$. Therefore, incorporating the second
covariant gauge fixing term into the action, we can write the generating
functional in the form (\ref{generatingfunction4})
\begin{equation}
Z = N \int {\cal D} A_{\mu}\,\Delta_{\rm FP} [A]
\left.\bar{\Delta}_{\rm FP} [A]\right|_{\tilde{n}\cdot x=\tau} 
\delta (n\cdot A) \,e^{iS},\label{generatingfunction5}
\end{equation}
where
\begin{equation}
S = S_{\rm inv} - \frac{1}{2\xi} \int d^{4}x\,\delta (\tilde{n}\cdot x
- \tau) \left(\partial\cdot A\right)^{2}.\label{secondfixing1}
\end{equation}
We note once again that $\Delta_{\rm FP} [A]$ is trivial in the
light-cone gauge, but the second Faddeev-Popov determinant leads to a
ghost action much like (\ref{ghostaction}) where the ghost action is
defined only for $\tilde{n}\cdot x = \tau$.

To define the propagator, we note that in the light-cone gauge
(\ref{lightconegauge1}), the Green's function of the theory described
by (\ref{secondfixing1}) would satisfy
\begin{equation}
\left(\delta^{\mu}_{\sigma} -
\frac{n^{\mu}\tilde{n}_{\sigma}}{n\cdot\tilde{n}}\right)
\left(\eta^{\sigma\lambda}
\Box - \partial^{\sigma}\partial^{\lambda} +
\frac{1}{\xi} \partial^{\sigma} \delta (\tilde{n}\cdot x-\tau)
\partial^{\lambda}\right) D_{\lambda\nu} (x,y) =
\left(\delta^{\mu}_{\nu} - \frac{n^{\mu} \tilde{n}_{\nu}}{n\cdot
  \tilde{n}}\right) 
\delta^{4} (x-y),\qquad n^{\mu} D_{\mu\nu} = 0 = D_{\mu\nu}
n^{\nu}.\label{greensfunction1}
\end{equation}
There are two linearly independent second rank symmetric projection
operators  which vanish when contracted with either $n^{\mu}$ or
$n^{\nu}$ (one was already given in (\ref{projection})),
\begin{eqnarray}
P_{\mu\nu} & = & \eta_{\mu\nu} - \frac{n_{\mu}\partial_{\nu} +
n_{\nu}\partial_{\mu}}{(n\cdot \partial)} +
\frac{\partial^{2}}{(n\cdot \partial)^{2}}\,n_{\mu}n_{\nu}\nonumber\\
Q_{\mu\nu} & = & \eta_{\mu\nu} - \frac{n_{\mu}\partial_{\nu} +
n_{\nu}\partial_{\mu}}{(n\cdot \partial)}.
\end{eqnarray}
It is easy to check that while both vanish when contracted with
$n^{\mu}$ or $n^{\nu}$, the first structure is in addition transverse
to $\partial^{\mu}$, namely
\begin{equation}
\partial^{\mu} P_{\mu\nu} = 0 = P_{\mu\nu} \partial^{\nu}.
\end{equation}

Thus, much like the case of the axial-type gauges, we can decompose the
propagator as
\begin{equation}
D_{\mu\nu} (x,y) = P_{\mu\nu} a (x,y) + \left(P_{\mu\nu} -
Q_{\mu\nu}\right) c (x,y) = P_{\mu\nu} a (x,y) + n_{\mu}n_{\nu} b
(x,y),\label{decomposition1}
\end{equation}
where we can think of $a (x,y), b (x,y)$ respectively as the
transverse and the longitudinal components of the
propagator. Substituting (\ref{decomposition1}) into
(\ref{greensfunction1}), we can derive the equations for the
coefficient functions to be
\begin{equation}
\Box a (x,y) = \delta^{4} (x-y),\qquad n\cdot \partial\left(1 -
\frac{1}{\xi} \delta (\tilde{n}\cdot x-\tau)\right) n\cdot \partial b
(x,y) = - \delta^{4} (x-y).\label{equations1}
\end{equation}
The first equation is easy to solve as in the case of the axial-type
gauges and leads to
\begin{equation}
a (x,y) = - \int \frac{\mathrm{d}^{4}k}{(2\pi)^{4}}\,\frac{e^{ik\cdot
(x-y)}}{k^{2}}.\label{transverse1}
\end{equation} 
The equation for the longitudinal component, on the other hand, is
quite different from that in (\ref{equations}) and this brings in new
features for the light-cone gauge. For example, we note that to be
able to solve (see (\ref{equations1}))
\begin{equation}
\left(1 - \frac{1}{\xi}\,\delta (\tilde{n}\cdot x - \tau)\right)
n\cdot 
\partial b (x,y) = - \frac{1}{n\cdot \partial}\,\delta^{4}
(x-y),\label{prescription}
\end{equation}
we need a prescription for $\frac{1}{n\cdot \partial}$. 
Let us represent
\begin{equation}
\frac{1}{n\cdot \partial}\,\delta \left(\tilde{n}\cdot x -
\tilde{n}\cdot y\right) = \Theta \left(\tilde{n}\cdot x -
\tilde{n}\cdot y\right),\label{theta}
\end{equation}
where $\Theta$ represents a generalized step function satisfying
\begin{equation}
n\cdot \partial\,\Theta \left(\tilde{n}\cdot x - \tilde{n}\cdot
y\right) = \delta \left(\tilde{n}\cdot x - \tilde{n}\cdot y\right).
\end{equation}
For example, we can have the naive representation of (\ref{theta}) as
the ordinary step function or an alternating step function if we
choose the  principal value prescription 
or the  Mandelstam-Leibbrandt prescription \cite{leibbrandt}. This
prescription  dependence, even after fixing the residual gauge
symmetry, is a new feature of the light-cone
gauge and reflects the fact that there is still some underlying
global invariance of the theory, such as the one discussed in the last
section, which leads to this arbitrariness. It is easy to check that
the global transformation of (\ref{globaltfn}) becomes only an
on-shell symmetry of the quadratic action because of the $\delta
(\tilde{n}\cdot 
x - \tau)$ term in the gauge fixing action. However, if we
generalize the transformation of (\ref{globaltfn}) as
\begin{equation}
\delta A_{\mu} = \zeta\,n_{\mu}\left(1- \frac{1}{\xi}\,\delta
(\tilde{n}\cdot x - \tau)\right) \partial\cdot A,\label{globaltfn2}
\end{equation}
this defines a global symmetry of the quadratic part of the action
(\ref{secondfixing1}) and
this is the origin of the arbitrariness (prescription dependence) in
the definition of the propagator. One needs to treat $\delta (x)$ and 
in particular $\delta (0)$ in
this derivation in a regularized manner from a definition such as 
\begin{equation}
\delta (x) = \lim_{\eta\rightarrow
0}\,\frac{1}{\sqrt{\pi}\eta}\,\exp\left(-\frac{x^{2}}{\eta^{2}}\right),
\label{delta0}
\end{equation}
with the understanding that the limit $\eta\rightarrow 0$ has to
be taken only at the end. 

It is well known that, in the light-cone gauge, prescriptions such as
the principal value for the unphysical poles lead to incorrect results
and the only consistent prescription that works correctly is the
Mandelstam-Leibbrandt prescription \cite{leibbrandt}. Therefore,
choosing this prescription, we can write
\begin{equation}
\Theta \left(\tilde{n}\cdot x - \tilde{n}\cdot y\right) =
\lim_{\eta\rightarrow 0}\;
\int \frac{\mathrm{d} (n\cdot k)}{2\pi i (n\cdot
\tilde{n})}\,\frac{e^{i(n\cdot k) 
\left(\tilde{n}\cdot x - \tilde{n}\cdot y\right)}}{(n\cdot k) -
i (\tilde{n}\cdot k) \eta} = \frac{1}{(n\cdot \tilde{n})}\,\epsilon
(\tilde{n}\cdot k)\,\theta \left(\tilde{n}\cdot k (\tilde{n}\cdot x -
\tilde{n}\cdot y)\right).\label{mandelstam}
\end{equation}
With such a prescription, it is straightforward to
check that the solution of (\ref{prescription}) or (\ref{equations1})
has the form
\begin{equation}
b (x,y) = - \epsilon (n\cdot \tilde{n})\left[\tilde{n}\cdot
x \Theta (\tilde{n}\cdot x-\tilde{n}\cdot y) +
\tilde{n}\cdot y \Theta (\tilde{n}\cdot y -
\tilde{n}\cdot x) - \frac{n\cdot \tilde{n}\Theta (\tau -\tilde{n}\cdot
  x) \Theta (\tau -\tilde{n}\cdot y)}{(\xi - \delta
(0))}\right] \delta^{2} (x_{\perp} - y_{\perp})\delta (n\cdot x -
n\cdot y),\label{longitudinal1}
\end{equation}
where the $\delta (0)$ term is necessary to impose the correct
boundary condition 
satisfied by the propagator and should be understood in a regularized
manner from a representation such as in (\ref{delta0}). The
longitudinal part of the propagator
is now uniquely determined from (\ref{longitudinal1}) and the
Mandelstam-Leibbrandt prescription (\ref{mandelstam}) then defines the
unphysical poles of the transverse part of the propagator as well. 

\section{Conclusion}

In this paper we have analyzed the question of residual gauge fixing
in the path integral approach in a systematic manner and have
determined the completely gauge fixed propagator in the
axial-type gauges as well as in the light-cone gauge. In both the
cases, the propagator can be defined without fixing the residual gauge
symmetry, but then one has to specify a prescription for handling the
unphysical poles in the propagator. In the case of the axial-type
gauges, the residual gauge fixing determines the propagator completely
without any problem of unphysical poles. In the light-cone gauge,
however, there is still a prescription dependence in the propagator
even after fixing the residual gauge symmetry. This reflects the
existence of a global invariance (\ref{globaltfn2}) of the quadratic
part of the theory which is the source of the arbitrariness in the
definition of the propagator. However, if we take the
Mandelstam-Leibbrandt  prescription (\ref{mandelstam}) which is the
conventional prescription in the light-cone gauge, it determines both
the  transverse as well the longitudinal parts of the propagator
completely. We note here that the completely gauge fixed
propagator in the path integral approach in axial type gauges as well
as the light-cone gauge in general continue
to be different from that in the Hamiltonian formalism (for
similar gauge fixing terms). Namely, in the Hamiltonian formalism the
propagator is doubly transverse (this is true in axial type gauges as
well) while the completely gauge fixed propagator in the path integral
approach is, in general, transverse with respect to $n^{\mu}$, but has
a longitudinal component with respect to $p^{\mu}$. 
Such a  difference is, in fact, natural and can be easily understood on
physical grounds as follows. In the Hamiltonian formalism, the
constraints can be set strongly equal to zero (after calculating the
Dirac brackets) thereby eliminating certain components of the
fields. This leads to analogues of nonlocal ``instantaneous Coulomb''
type interaction terms in the Hamiltonian. On the other hand, in the path
integral formalism, one does not explicitly eliminate components of
the fields and correspondingly, such interactions arise only through the
exchange of longitudinal gluons (longitudinal components of the gauge
propagator) and the ghosts. The longitudinal components of the
propagator are, therefore, absolutely essential in the path integral
approach together with the ghost terms (ghosts are necessary to cancel out
any dependence of physical quantities on $\tau$)
as has been stressed within the context of the calculation of the
Wilson line in the temporal gauge \cite{rossi}.

In summary, we note that in this paper, our goal has been to compare
the form of the path integral propagator with that obtained from a
Hamiltonian analysis. To that end, we have chosen the residual gauge
fixing to be the covariant gauge in a manner completely parallel with
the Hamiltonian analysis. As we have shown, with this choice in the
light-cone gauge, there is a residual global symmetry of the free
action (both in the Hamiltonian as well as the path integral
formalisms). In fact, there is as well an Abelian local gauge invariance
\begin{equation}
A_{\mu} (x) \rightarrow A_{\mu} (x) + \partial_{\mu} \hat{\theta}
(n\cdot x),
\end{equation}
of the free action where $\hat{\theta} (n\cdot x)$ depends only on
$n\cdot x$. The presence of these residual symmetries leads to the
prescription dependence in the propagator, both in the Hamiltonian as
well as in the path integral formalisms. The prescription dependence can
be eliminated in the path integral formalism much like in the axial
type gauges if one chooses a residual gauge fixing term which leaves
no further global/local invariance in the free action. The choice of
such a residual gauge in the context of the light-cone gauge is
presently under study and the results will be reported in future.

\vskip .7cm

\noindent{\bf Acknowledgment:}
\medskip

We would like to thank Prof. J. C. Taylor for several helpful remarks.
Two of us (A.D. and S.P.) would like to thank the Rockefeller
Foundation for a fellowship for residency and Ms. Gianna Celli for the
kind hospitality at
the Bellagio Study Center where part of this work was done. This work
was supported in part by the US DOE Grant number DE-FG 02-91ER40685
and by CNPq as well as by FAPESP, Brazil.

\appendix

\section{Propagator and the inverse two point function in the
light-cone gauge}

In this appendix, we will show very briefly that the propagator in the
naive light-cone gauge in the path integral approach corresponds to
the inverse of the two point
function given in (\ref{pathintegralinverse}). Let us consider the
generating function (\ref{generatingfunction}) in the presence of
sources.
\begin{equation}
Z[J^{\mu},J] = N \int {\cal D}A_{\mu}\,{\cal D}F\,e^{i(S+ S_{\rm
source})}, \quad S =
S_{\rm inv} - \int \mathrm{d}^{4}x {\rm Tr} F n\cdot A,\quad S_{\rm
source} = \int \mathrm{d}^{4}x\,{\rm Tr}\left(J^{\mu}A_{\mu} +
JF\right).\label{app1}
\end{equation}
Here we have absorbed the non-dynamical Faddeev-Popov determinant into
the normalization constant and have exponentiated the delta function
constraint with the help of an auxiliary field.

It is straightforward to determine the classical fields from the form
of the action $S$ in (\ref{app1}) and they take the forms
\begin{eqnarray}
F^{c} & = & \frac{1}{n\cdot
\partial}\,\partial_{\mu}J^{\mu},\nonumber\\
A_{\mu}^{c} & = & \frac{1}{\Box}\left(g_{\mu\nu} -
\frac{n_{\mu}\partial_{\nu} + n_{\nu}\partial_{\mu}}{n\cdot
\partial}\right)J^{\nu} + \frac{1}{n\cdot \partial}\,\partial_{\mu}
J = \left(\Gamma^{\rm (PI)}\right)^{-1}_{\mu\nu} (n,\partial) J^{\nu}
+ \frac{1}{n\cdot \partial}\,\partial_{\mu} J,\label{app2}
\end{eqnarray}
where $\left(\Gamma^{\rm (PI)}\right)^{-1}_{\mu\nu} (n,\partial)$ is
the inverse of the two point function given in
(\ref{pathintegralinverse}) in the coordinate space.
Shifting the fields in the generating function (\ref{app1}) by
$$
A_{\mu}\rightarrow A_{\mu} + A_{\mu}^{c},\qquad F\rightarrow F +
F^{c},
$$
we obtain
\begin{equation}
Z[J^{\mu}, J] = e^{i\int \mathrm{d}^{4}x\,{\rm Tr}\left[\frac{1}{2}
J^{\mu}\left(\Gamma^{\rm (PI)}\right)^{-1}_{\mu\nu} (n,\partial) J^{\nu}
+ J \frac{1}{n\cdot \partial} \partial_{\mu}J^{\mu}\right]}\;N\int
{\cal D}A_{\mu}\,{\cal D}F\,e^{iS}.
\end{equation}
This determines that the propagator for the gauge
field in the naive light-cone gauge in the path integral approach has
the form 
\begin{equation}
D_{\mu\nu}^{\rm (PI)} = - \left.\frac{1}{Z}\,\frac{\delta^{2}Z}{\delta
J^{\mu}\delta J^{\nu}}\right|_{J^{\mu},J=0} = \left(\Gamma^{\rm
(PI)}\right)^{-1}_{\mu\nu} (n,\partial),
\end{equation}
as claimed in the text.

\section{Some useful formulae}

In this appendix, we compile some formulae that are quite useful in
dealing with light-cone variables. First, let us assume that the
light-like vectors $n^{\mu},\tilde{n}^{\mu}$ have vanishing components
along $i=1,2$ (transverse) directions. In that case, we can introduce
a new set of coordinates
\begin{equation}
\bar{x}^{\alpha} = \left(n\cdot x, x^{i}, \tilde{n}\cdot x\right) =
L^{\alpha}_{\;\mu} x^{\mu},
\end{equation}
where $x^{\mu}$ represents the usual Minkowski coordinates and
\begin{equation}
L^{\alpha}_{\;\mu} = \left(\begin{array}{cccc}
n^{0} & 0 & 0 & - n^{3}\\
0 & 1 & 0 & 0\\
0 & 0 & 1 & 0\\
\tilde{n}^{0} & 0 & 0 & - \tilde{n}^{3}
\end{array}\right).
\end{equation}
The metric tensors for the new coordinates take the forms
\begin{equation}
\bar{g}^{\alpha\beta} = \left(\begin{array}{crrc}
0 & 0 & 0 & n\cdot \tilde{n}\\
0 & -1 & 0 & 0\\
0 & 0 & -1 & 0\\
n\cdot \tilde{n} & 0 & 0 & 0
\end{array}\right),\quad \bar{g}_{\alpha\beta} =
\left(\begin{array}{crrc} 
0 & 0 & 0 & \frac{1}{n\cdot \tilde{n}}\\
0 & -1 & 0 & 0\\
0 & 0 & -1 & 0\\
\frac{1}{n\cdot \tilde{n}} & 0 & 0 & 0
\end{array}\right).
\end{equation}
The integration measure correspondingly can be written as
\begin{equation}
\int \mathrm{d}^{4}x = \frac{1}{|n\cdot \tilde{n}|} \int
\mathrm{d}^{4}\bar{x}.
\end{equation}
Furthermore, the delta functions would transform as
\begin{equation}
\delta^{4} (x) = |n\cdot \tilde{n}| \delta^{4} (\bar{x}).
\end{equation}
These are some of the formulae that have been used in the derivations
in the paper.


\bigskip

\begin{thebibliography}{100}

\bibitem{dirac} P. A. M. Dirac, Reviews of Modern
Physics {\bf 21}, 392 (1949); {\em Lectures in Quantum Mechanics}
(Benjamin, New York, 1964); A. J. Hanson, T. Regge and C. Teitelboim,
{\em Constrained Hamiltonian Systems} (Academia Nazionale dei Lincei,
Rome, 1976).

\bibitem{brodsky} There are numerous papers on the subject. We are
going to refer the readers to only a few well known review articles
where further references can be found. S. J. Brodsky, H. C. Pauli and
S. S. Pinsky, Physics Reports {\bf 301}, 299 (1998); K. Yamawaki,
\lq\lq Zero mode problem on the light-front'', hep-th/9802037;
T. Heinzl, \lq\lq Light-cone quantization: Foundations and
applications'', Lecture Notes in Physics {\bf 572}, 55 (2001).

\bibitem{das} V. S. Alves, A. Das and S. Perez, Physical Review D
{\bf 66}, 125008 (2002); A. Das, hep-th/0310247, Proceedings of {\em
Hadrons and Beyond (LC03)}, Durham, England, 2003.

\bibitem{weldon} H. A. Weldon, Physical Review D {\bf 67},
085027  (2003).

\bibitem{das1} A. Das and X. Zhou, Physical Review D {\bf 68},
065017 (2003).

\bibitem{brodsky1} P. P. Srivastava and S. Brodsky, Phys. Rev. D {\bf
64}, 045006 (2001); Phys. Rev. D {\bf 66}, 045019 (2002).

\bibitem{perez} A. Das and S. Perez, Physical Review  D {\bf 70},
  065006 (2004).

\bibitem{suzuki} A. Suzuki and J. H. O. Sales, Nuc. Phys. A {\bf 725},
139 (2003); Mod. Phys. Lett. A {\bf 19}, 1925 (2004).

\bibitem{rossi} J. P. Leroy, J. Micheli and G. C. Rossi, Z. Phys. C
{\bf 36}, 305 (1987).

\bibitem{leibbrandt} G. Leibbrandt, 
Reviews of Modern Physics {\bf 59}, 1067 (1987).

\bibitem{faddeev} L. D. Faddeev and V. N. Popov, Phys. Lett. B {\bf
25}, 29 (1967); FERMILAB-PUB-72-057-T, NAL-THY-57, May 1972.

\bibitem{frenkel} J. Frenkel, Phys. Rev. D {\bf 13}, 2325 (1976).

\bibitem{BRS} C. Becchi, A. Rouet and R. Stora, Ann. Phys. {\bf 98},
287 (1976).

\bibitem{girotti} G. C. Rossi and M. Testa, Nuc. Phys. B {\bf 163},
  109 (1980); Phys. Rev. D {\bf 29}, 2997 (1984); S. Caracciolo,
  G. Curci and P. Menotti, Phys. Lett. B {\bf 113}, 311 (1982);
  H. P. Dahmen, B. Scholz and F. Steiner, Phys. Lett. {\bf
117B}, 339 (1982); A. Burnel, Phys. Rev. D {\bf 27}, 1830 (1983);
  H. O. Girotti and H. J. Rothe,  Z. Phys. C {\bf 27},
559 (1985); P. V. Landshoff, Phys. Lett. B {\bf 169}, 69 (1986).

\bibitem{thooft} G. 't Hooft, {\em Diagrammar}, CERN Pub-73-9, 1973 ;
{\em Under the Spell of the Gauge Principle}, World Scientific,
Singapore, 1994.

\bibitem{bassetto} A. Bassetto, M. Dalbosco, I. Lazzizzera and
  R. Soldati, Phys. Rev. D {\bf 31}, 2012 (1985); M. Dalbosco,
  Phys. Lett. B {\bf 163}, 181 (1985); Phys. Lett. B {\bf 180}, 121
  (1986). 


\end{thebibliography}
\end{document}